\documentclass[12pt]{iopart}

\usepackage{iopams}  
\usepackage{graphicx}
\usepackage{bm}
\usepackage{cite}
\usepackage{color}
\usepackage{url}
\usepackage{hyperref}
\usepackage{bbold}

\newcommand*{\vek}[1]{{\ensuremath{\bm{\mathrm{#1}}}}}
\newcommand* {\kk}{\vek{k}}

\newcommand* {\ket}[1]{\ensuremath{| {#1} \rangle}}
\newcommand* {\braket}[1]{\ensuremath{\langle {#1} \rangle}}
\newcommand{\floor}[1]{\left\lfloor #1 \right\rfloor}
\providecommand{\brac}[1]{\left(#1\right)} 
\providecommand{\Brac}[1]{\left[#1\right]} 

\graphicspath{{.}{./EPS/}{./Figures/}}

\begin{document}

\title{Manipulating topological-insulator properties using quantum confinement}

\author{M Kotulla, U Z\"ulicke}

\address{School of Chemical and Physical Sciences and MacDiarmid Institute
for Advanced Materials and Nanotechnology, Victoria University of Wellington,
PO Box 600, Wellington 6140, New Zealand}

\ead{uli.zuelicke@vuw.ac.nz}

\vspace{10pt}
\begin{indented}
\item[]\today
\end{indented}

\begin{abstract}

Recent discoveries have spurred the theoretical prediction and experimental
realization of novel materials that have topological properties arising from
band inversion. Such topological insulators are insulating in the bulk but
have conductive surface or edge states. Topological materials show various
unusual physical properties and are surmised to enable the creation of exotic
Majorana-fermion quasiparticles. How the signatures of topological behavior
evolve when the system size is reduced is interesting from both a fundamental
and an application-oriented point of view, as such understanding may form the
basis for tailoring systems to be in specific topological phases. This work
considers the specific case of quantum-well confinement defining
two-dimensional layers. Based on the effective-Hamiltonian description of bulk
topological insulators, and using a harmonic-oscillator potential as an example
for a softer-than-hard-wall confinement, we have studied the interplay of band
inversion and size quantization. Our model system provides a useful platform
for systematic study of the transition between the normal and topological
phases, including the development of band inversion and the formation of
massless-Dirac-fermion surface states. The effects of bare size quantization,
two-dimensional-subband mixing, and electron-hole asymmetry are disentangled
and their respective physical consequences elucidated.

\end{abstract}

%
%
\submitto{\NJP}
%
\maketitle

\section{Introduction}

Topological insulators~\cite{has10,qi11,has15} (TIs) have emerged as a new
materials class that provides a testing ground for exploring ground-breaking
new ideas~\cite{chi16,lud16} about the properties of condensed matter.
Realizations are possible in electronically two-dimensional~\cite{kan05,ber06,
kon07,liu08,kne11} (2D) or three-dimensional~\cite{has11} (3D) systems, and
simulations using cold atoms subject to artificial gauge fields in optical
lattices have been explored~\cite{sta09,sta10,ber10,gol10,ber11}. Quantum
confinement was recognized early on as a mechanism for causing a transition
between topological and normal phases~\cite{ber06,kon07,liu10a}, as well as a
tool to manipulate the massless-Dirac-like surface (edge) states~\cite{lin09,
liu10a,lu10,sha10,zha12,zho08} of a 3D (2D) TI. Recently, the interplay between
topological properties and quantum confinement has been studied as a way to
disentangle various phenomena typically associated with topological materials
(protected surface states, band inversion, a finite gap) in a controlled
fashion~\cite{sal16}.

We utilize an effective-model description of quantum wells made from TI
materials to map changes in the size of the fundamental gap, the degree of band
inversion, and the hybridization of surface states in a detailed way. As we
augment the generic $\mathbf{k}\cdot\mathbf{p}$ Hamiltonian~\cite{zha09,liu10}
for 3D TIs with a harmonic confinement potential, our work is complementary to
previous studies~\cite{lin09,liu10a,lu10,sha10,zha12} where a hard-wall
confinement was considered. Further motivation to investigate influences of the
confinement strength on 3D-TI properties is provided by the observation that 2D
TIs (realized in HgTe quantum wells or simulated by cold atoms, respectively)
subjected to soft confinement experience a proliferation of edge modes that
modifies topological properties~\cite{sta09, sta10,buc12,wan17}. Our results
serve to differentiate truly universal behavior from properties that depend on
the material and/or the dimensionality. We identify Bi$_2$Te$_3$ and
Sb$_2$Te$_3$ as materials where effects due to size quantization are
sufficiently prominent to be accessible experimentally, whereas Bi$_2$Se$_3$ is
typically not in a confinement-dominated regime. Strong mixing of 2D subbands
occurs for all layered materials in the limit of thicker samples, resulting in
the formation of the 2D-Dirac-like surface states. Electron-hole asymmetry can
partly obscure the observability of topological effects. Unravelling the
complexity of these various influences enables us to provide insights to guide
future experimental study of confined topological insulators.

The remainder of this paper is organised as follows. In 
section~\ref{sec:model}, our effective-model description of a
quantum-well-confined 3D TI is introduced and relevant notation established.
For instruction and greater clarity, basic ramifications of bare subband
quantization and confinement-induced subband mixing are discussed for the
electron-hole-symmetric case in sections~\ref{sec:subbands} and
\ref{sec:bandinv}, before presenting results for the realistic situation with
typically sizable electron-hole asymmetry in section~\ref{sec:ehAS}. Signatures
of topological behavior in the subband dispersions (section~\ref{sec:subbands})
and in bound-state properties such as the band-inversion-related pseudospin and
the surface-state hybridization (section~\ref{sec:bandinv}) are identified,
also with respect to their observability in real (electron-hole-asymmetric)
systems. Our results are summarized in section~\ref{sec:concl} and further
illustrated by interactive simulations that are provided as supplementary data.
Some more mathematical details about the methods used for calculating
3D-TI-layer subband dispersions and eigenstates are given in the appendix.  

\section{Model for quantum-confined 3D topological insulators}
\label{sec:model}

Our starting point is the effective low-energy model Hamiltonian~\cite{zha09,
liu10} describing the 3D-TI material family $X_2Y_3$, with $X\in\{\mathrm{Bi},
\mathrm{Sb}\}$ and $Y\in\{\mathrm{Se}, \mathrm{Te}\}$. We use the
representation where basis states are ordered as $\ket{P1^+_-, \frac{1}{2}}$,
$i \ket{P2^-_+,-\frac{1}{2}}$, $\ket{P1^+_-,-\frac{1}{2}}$, $- i \ket{P2^-_+,
\frac{1}{2}}$ and write $H = H_0 + H_\parallel + H_\epsilon$, with
\numparts
\begin{eqnarray}
H_0(k_z, \kk_\perp) &=& \left( \begin{array}{cc} h_0(k_z) + h_\perp(\kk_\perp)
& \mathbf{0} \\ \mathbf{0} & h_0(k_z) + h_\perp^\ast(\kk_\perp) \end{array}
\right) \quad , \\
H_\parallel(k_z) &=& \left( \begin{array}{cc} \mathbf{0} & i \,  B_0 \, k_z\,
\tau_y \\ -i \,  B_0 \, k_z\, \tau_y & \mathbf{0} \end{array} \right) \quad , 
\\ H_\epsilon(k_z,\kk_\perp) &=& \left( \begin{array}{cc} \epsilon(k_z,
\kk_\perp )\, \tau_0 & \mathbf{0} \\ \mathbf{0} & \epsilon(k_z, \kk_\perp)\,
\tau_0 \end{array} \right) \quad . \label{eq:phAsymmH}
\end{eqnarray}
\endnumparts
The asterisk indicates complex conjugation. We introduced $\tau_{x, y,z}$ and
$\tau_0$ to denote, respectively, the Pauli matrices and the identity matrix
acting in the pseudo-spin-$1/2$ subspaces spanned by the conduction and
valence-band-edge states $\ket{P1^+_-,\sigma}$, $\sigma i \ket{P2^-_+,
-\sigma}$ for fixed $\sigma$, and used the abbreviations
\numparts
\begin{eqnarray}
h_0(k_z) &=& \left( M_0 + M_1\, k_z^2 \right) \tau_z \quad , \\
h_\perp(\kk_\perp) &=& M_2\, \kk_\perp^2 \, \tau_z + A_0 \left( k_x\, \tau_x
+ k_y\, \tau_y \right) \quad , \label{eq:inPlaneH} \\
\epsilon(k_z, \kk_\perp) & =& C_0 + C_1\, k_z^2 + C_2\, \kk_\perp^2 \quad .
\end{eqnarray}
\endnumparts
Somewhat differing values for the band-structure parameters $M_j$, $A_0$, $B_0$
and $C_j$ applicable to currently available 3D-TI materials have been quoted in
the literature~\cite{zha09,liu10,sha10,orl15,nec16}. To be specific, we use
band-structure parameters from Table~I of Ref.~\cite{nec16}. Although the
absolute values differ, basic materials-related trends found in our work
generally emerge the same way when previously reported parameter sets are used,
e.g., those from Table~IV of Ref.~\cite{liu10}. The fact that $M_0 < 0$
embodies the band inversion responsible for the material's topological
properties.

The partition of $H$ into its three parts $H_0$, $H_\parallel$ and $H_\epsilon$
is motivated by the intention to discuss the interplay of various TI
band-structure features with a confining potential in $z$ direction. We choose
the latter to be of the harmonic-oscillator form given by
\numparts
\begin{eqnarray}
H_V(z) &=& \left( \begin{array}{cc} V(z) \, \tau_z & \mathbf{0} \\ \mathbf{0}
& V(z) \, \tau_z \end{array} \right) \quad , \\ V(z) &=& \,\, \frac{\hbar^2}{4
M_1} \, \Omega^2\, z^2 \quad .
\end{eqnarray}
\endnumparts
Although it will generally not be a fully realistic approximation for the
electronic confinement in thin layers of TI materials, the harmonic-oscillator
potential enables an illuminating theoretical treatment of quantum-confinement
effects that yields universally applicable conclusions. Furthermore,
investigating the features that are specific to a smooth confinement, in
contrast to the typically considered abrupt hard-wall~\cite{zho08,lin09,liu10a,
lu10,sha10} or infinitely steep inverted-mass~\cite{zha12} type, can yield
interesting new insights into basic properties of topological
protection~\cite{wan17,buc12}.

To start with, we consider the basic confinement problem embodied in $H_0
(-i\partial_z, \vek{0}) + H_V(z)$, which can be reduced to a Schr\"odinger
equation for the individual blocks representing pseudo-spin-1/2 subspaces that
is given by
\begin{equation}
\left[ h_0( -i\partial_z ) + V(z) \,\tau_z \right] \psi_\vek{0}(z) =
E_\vek{0} \,\psi_\vek{0}(z) \quad .
\end{equation}
It has the energy eigenvalues $E_{n\vek{0}}^{(\tau)} = \tau [ M_0 + \hbar
\Omega (n + 1/2)]$ with $n = 0, 1, \dots$, and the corresponding eigenstates
can be written as $\psi_{n\vek{0}}^{(\tau)}(z) = \phi_n (z/l_\Omega) \otimes
\ket{\tau}_\mathrm{p}$, where $\phi_n(\cdot)$ are familiar harmonic-oscillator
eigenfunctions, and $\ket{\tau}_\mathrm{p}$ denote the eigenstates of $\tau_z$
with eigenvalues $\tau = \pm 1$ that are associated with the opposite-parity
basis states. This establishes the characteristic energy scale $\hbar\Omega$
for size quantization and $l_\Omega\equiv \sqrt{2 M_1/(\hbar\Omega)}$ as a
measure for system size in the confined dimension. Relating these scales to
those relevant for real samples enables useful comparisons and detailed
predictions based on our model. The form of the energy eigenvalues $E_{n 
\vek{0}}^{(\tau)}$ also foreshadows how the positive size-quantization energy
competes with band inversion arising from a negative $M_0$.

\section{Emergence of size-quantized subband structure and surface states}
\label{sec:subbands}

We now focus on deriving a general description for the in-plane motion of
electrons in the confined 3D TI. To be able to express our results in terms of
a minimal set of materials-dependent quantities, we introduce the momentum and
energy units $q_\perp = A_0/M_2$ and $E_\perp = A_0 q_\perp\equiv A_0^2/M_2$.
All other relevant energy scales of our system of interest are then also
usefully measured in terms of $E_\perp$, yielding the parameters
\begin{equation}\label{eq:parameters}
\gamma_{M_0} = \frac{M_0}{E_\perp} \quad , \quad \gamma_\Omega = \frac{\hbar
\Omega}{E_\perp} \equiv \frac{2 M_1}{M_2}\, (q_\perp l_\Omega)^{-2} \quad ,
\quad \gamma_\parallel = \frac{B_0^2}{A_0^2}\, \frac{M_2}{2 M_1} \quad .
\end{equation}
Confined motion in $z$ direction and free motion in the $xy$ plane are
simultaneously contained in the Hamiltonian $H_0(-i\partial_z, \kk_\perp) +
H_V(z)$, which is block-diagonal. Focusing on the pseudo-spin-1/2 subspace
spanned by the basis states $\ket{P1^+_-,\frac{1}{2}}$, $i \ket{P2^-_+,-
\frac{1}{2}}$ yields the Schr\"odinger equation
\begin{equation}\label{eq:inplane}
\left[ h_0(-i\partial_z) + h_\perp (\kk_\perp) + V(z) \,\tau_z \right]
\psi_{\kk_\perp}(z) = E_{\kk_\perp} \,\psi_{\kk_\perp}(z) \quad .
\end{equation}
Looking for its eigenstates in terms of the superposition $\psi_{n \kk_\perp}
(z) =\sum_{\tau =\pm} a_{n \kk_\perp}^{(\tau)} \, \psi_{n\vek{0}}^{(\tau)}(z)$
transforms (\ref{eq:inplane}) into an effective 2D Dirac equation for the
subband dispersions,
\numparts
\begin{eqnarray}\label{eq:2D-Dirac}
&& \left[ \frac{\Delta_{n \kk_\perp}}{E_\perp} \, \tau_z + \frac{k_x}{q_\perp}
\, \tau_x + \frac{k_y}{q_\perp}\, \tau_y \right] \left( \begin{array}{c} a_{n
\kk_\perp}^{(+)} \\[0.1cm] a_{n \kk_\perp}^{(-)} \end{array} \right)
= \frac{E_{n \kk_\perp}}{E_\perp} \left( \begin{array}{c} a_{n\kk_\perp}^{(+)}
\\[0.1cm] a_{n \kk_\perp}^{(-)} \end{array} \right) \quad , \\[0.2cm]
&& \frac{\Delta_{n \kk_\perp}}{E_\perp} = \gamma_{M_0} + \gamma_\Omega \left(
n + \frac{1}{2} \right) + \frac{\kk_\perp^2}{q_\perp^2} \quad ,
\label{eq:topGap}
\end{eqnarray}
\endnumparts
which has a gap parameter $\Delta_{n \kk_\perp}$ that depends both on the
confinement and on the wave vector $\kk_\perp$ for in-plane motion. In
particular, only the subbands with $\Delta_{n \vek{0}} < 0$ still have inverted
character close to their band edges, which implies $n\le n_\mathrm{c}$ where
\begin{equation}
n_\mathrm{c} = \floor{-\frac{\gamma_{M_0}}{\gamma_\Omega} - \frac{1}{2}}
\quad .
\end{equation}
Thus similarly to the case of 2D TIs realized in semiconductor 
heterostructures~\cite{kon07,kne11}, quantum confinement drives transitions
between topological and normal phases of individual subbands~\cite{liu10a}.
For $\gamma_\Omega > -2\, \gamma_{M_0}$, all subbands are in the normal phase.

\begin{figure}
\begin{flushright}
\includegraphics[width=0.4\columnwidth]{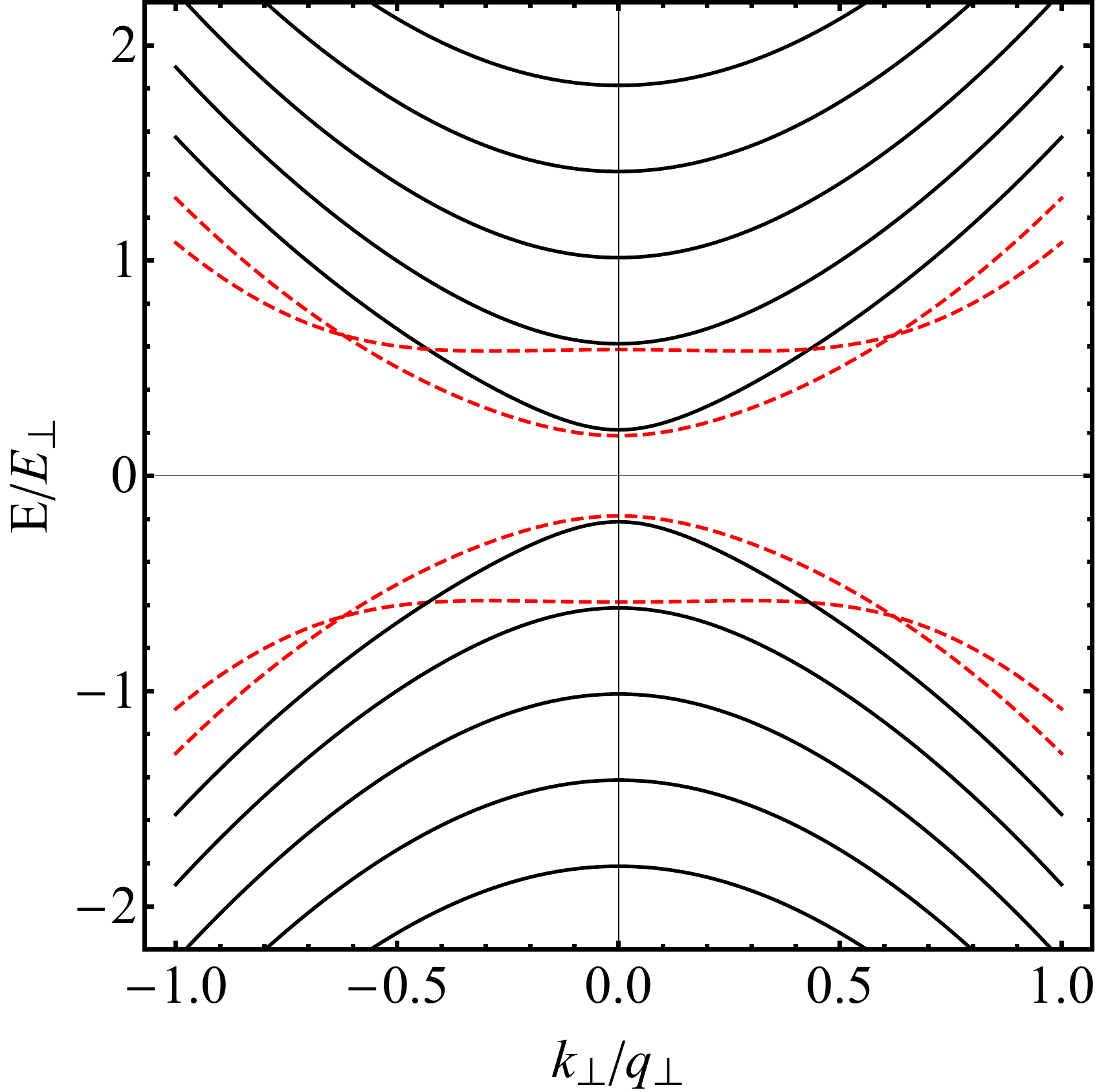}
\hspace{0.5cm}
\includegraphics[width=0.4\columnwidth]{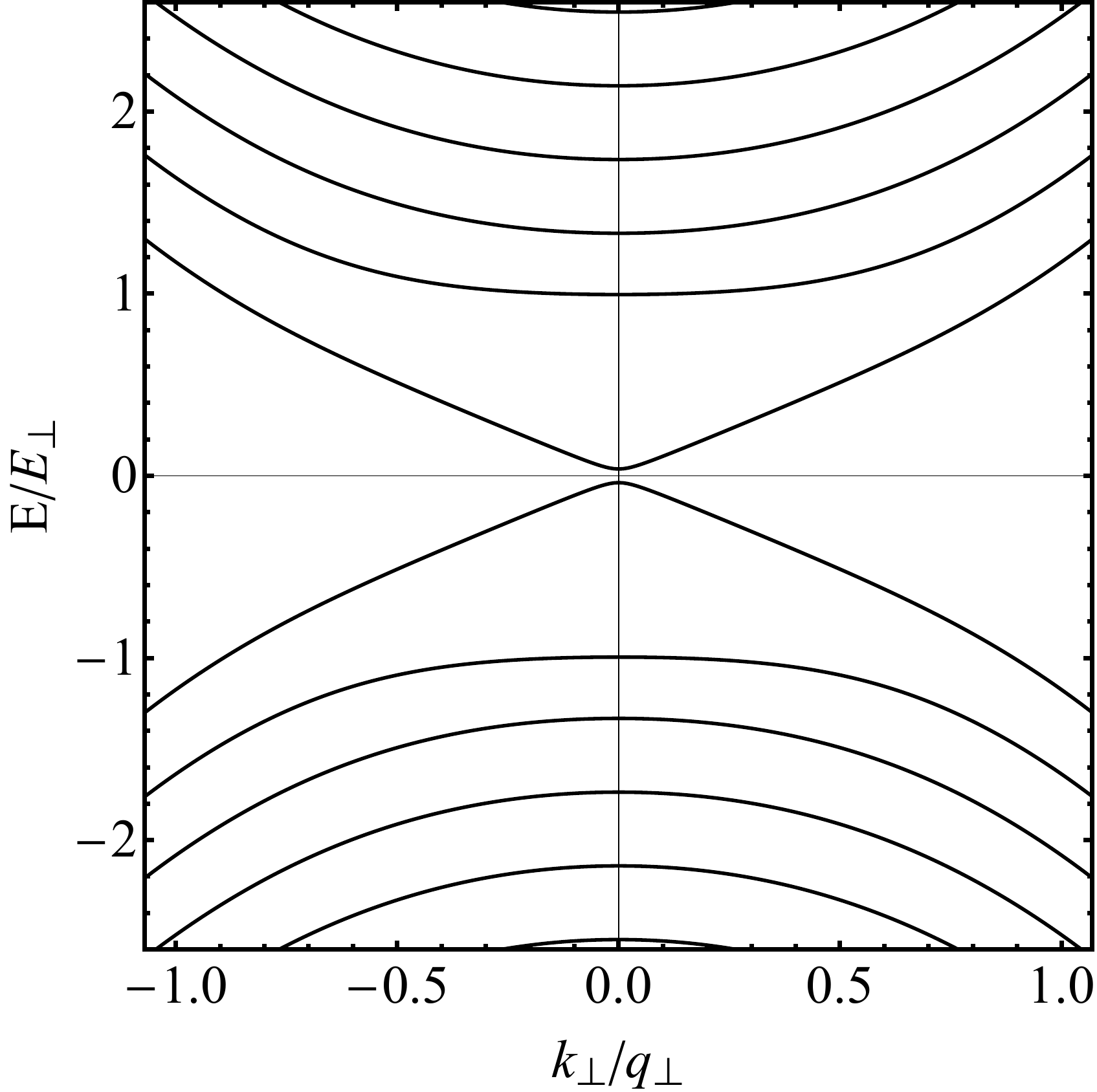}
\caption{Subband dispersions of the harmonically confined 3D topological
insulator. Left panel: Subbands obtained when both electron-hole asymmetry
($H_\epsilon$) and the $H_\parallel$-induced band mixing between the bare
harmonic-oscillator subbands are neglected. Dashed curves indicate inverted
bands. Right panel: Subbands calculated with only electron-hole asymmetry
($H_\epsilon$) neglected. In contrast to the case in the left panel, the
lowest subband exhibits the expected dispersion of weakly hybridized
massless-Dirac-like surface states. Parameters used in the calculations are
$\gamma_{M_0} = -0.785$, $\gamma_\parallel = 2.34$ (both correspond to
Bi$_2$Se$_3$ \cite{nec16}), and $\gamma_\Omega = 0.400$.}
\label{fig:subbands}
\end{flushright}
\end{figure}

Diagonalization of (\ref{eq:2D-Dirac}) is straightforward and yields the
eigenvalues and eigenstate spinor components
\numparts
\begin{eqnarray}\label{eq:perpEVs}
E_{n \kk_\perp \alpha} &=& \alpha\, E_\perp \,\, \sqrt{ \left( \frac{\Delta_{n
\kk_\perp}}{E_\perp} \right)^2 + \left( \frac{k_\perp}{q_\perp}\right)^2 }
\quad , \\[0.1cm]
a_{n \kk_\perp \alpha}^{(\tau)} &=& \alpha^{\frac{1-\tau}{2}} \,\, \sqrt{
\frac{k_x - \tau\, i\, k_y}{k_\perp}} \,\, \sqrt{\frac{E_{n \kk_\perp \alpha}
+ \tau\, \Delta_{n \kk_\perp}}{2\, E_{n \kk_\perp \alpha}}} \quad ,
\end{eqnarray}
\endnumparts
where $\alpha = \pm 1$ distinguishes positive-energy (conduction) and
negative-energy (valence) subbands. The energy spectrum exhibits the axial
symmetry of in-plane motion through the dependence of $E_{n \kk_\perp \alpha}$
on $k_\perp \equiv \sqrt{k_x^2 + k_y^2}$. Due to the block-diagonal form of the
Hamiltonian $H_0(-i\partial_z, \kk_\perp) + H_V(z)$, its eigenvalues are those
given by (\ref{eq:perpEVs}); each of them doubly degenerate with corresponding
eigenstates $\psi_{n\kk_\perp\alpha}(z) \otimes \ket{+}_\mathrm{s}$ and
$\psi^\ast_{n\kk_\perp\alpha}(z) \otimes \ket{-}_\mathrm{s}$. Here the
$\ket{\sigma}_\mathrm{s}$ denote eigenstates of $\sigma_z$ that distinguish the
conduction and valence-band-edge subspaces spanned by $\{ \ket{P1^+_-,\sigma},
\sigma i \ket{P2^-_+,-\sigma}\}$ with opposite $\sigma=\pm 1$. The left panel
in Fig.~\ref{fig:subbands} shows the dispersions obtained for a situation with
$n_\mathrm{c} = 1$.

While the oscillator subbands obtained from diagonalizing $H_0(-i\partial_z,
\kk_\perp) + H_V(z)$ illustrate how confinement affects topological band
inversion on a basic level, they can only yield a correct description when
$H_\parallel(-i\partial_z)$ is a small perturbation, i.e., in the
confinement-dominated limit where $\gamma_\parallel \ll \gamma_\Omega$. In
Bi$_2$Se$_3$, this is not the typical situation as $\gamma_\parallel = 2.34$
is larger than any reasonably possible value for $\gamma_\Omega$,\footnote{An
order-of-magnitude estimate based on the na{\"\i}ve identification $\sqrt{2}\, 
l_\Omega\equiv 1$~nm (approx.\ value of the Bi$_2$Se$_3$ quintuple-layer
thickness~\cite{zha09b}) implies $\gamma_\Omega \le 0.6$ (using materials
parameters from Ref.~\cite{nec16}).} and the inclusion of $H_\parallel(-i
\partial_z)$ significantly affects the subband dispersions. Formally, the
coupling of individual oscillator subbands by $H_\parallel(-i \partial_z)$ as
the 3D limit is approached has many similarities with the way models for 3D TIs
have been constructed based on transversely coupled 2D TIs~\cite{kob15}. Using
the eigenstates of $H_0(-i\partial_z, \kk_\perp) + H_V(z)$ as a basis, we are
able to efficiently diagonalize the full Hamiltonian $H_0 (-i\partial_z,
\kk_\perp) + H_\parallel(-i\partial_z) + H_V(z)$. See the Appendix for
mathematical details and the right panel of Fig.~\ref{fig:subbands} for
illustrative results. Most crucially, the band mixing induced by $H_\parallel
(-i\partial_z)$ conspires to create the surface-state dispersion expected for a
confined 3D TI in the large-width limit~\cite{lin09,liu10a,lu10,sha10,zha12}.
In contrast to Bi$_2$Se$_3$, the parameter $\gamma_\parallel$ is quite small
for Bi$_2$Te$_3$ ($0.484$) and Sb$_2$Te$_3$ ($0.117$), as compared with the
nominal maximum values of $\gamma_\Omega$ ($4$ and $3$) derived from equating
$\sqrt{2}\, l_\Omega$ with their quintuple-layer width ($\approx 1$~nm). Hence,
these materials should lend themselves to a more detailed exploration of the
transition between the 3D-bulk and confinement-dominated regimes. The
interactive simulations provided in the supplementary data for this article
show the evolution of subband structure in confined TIs made from Bi$_2$Se$_3$,
Bi$_2$Te$_3$, and Sb$_2$Te$_3$.

\begin{figure}
\begin{flushright}
\includegraphics[width=0.42\columnwidth]{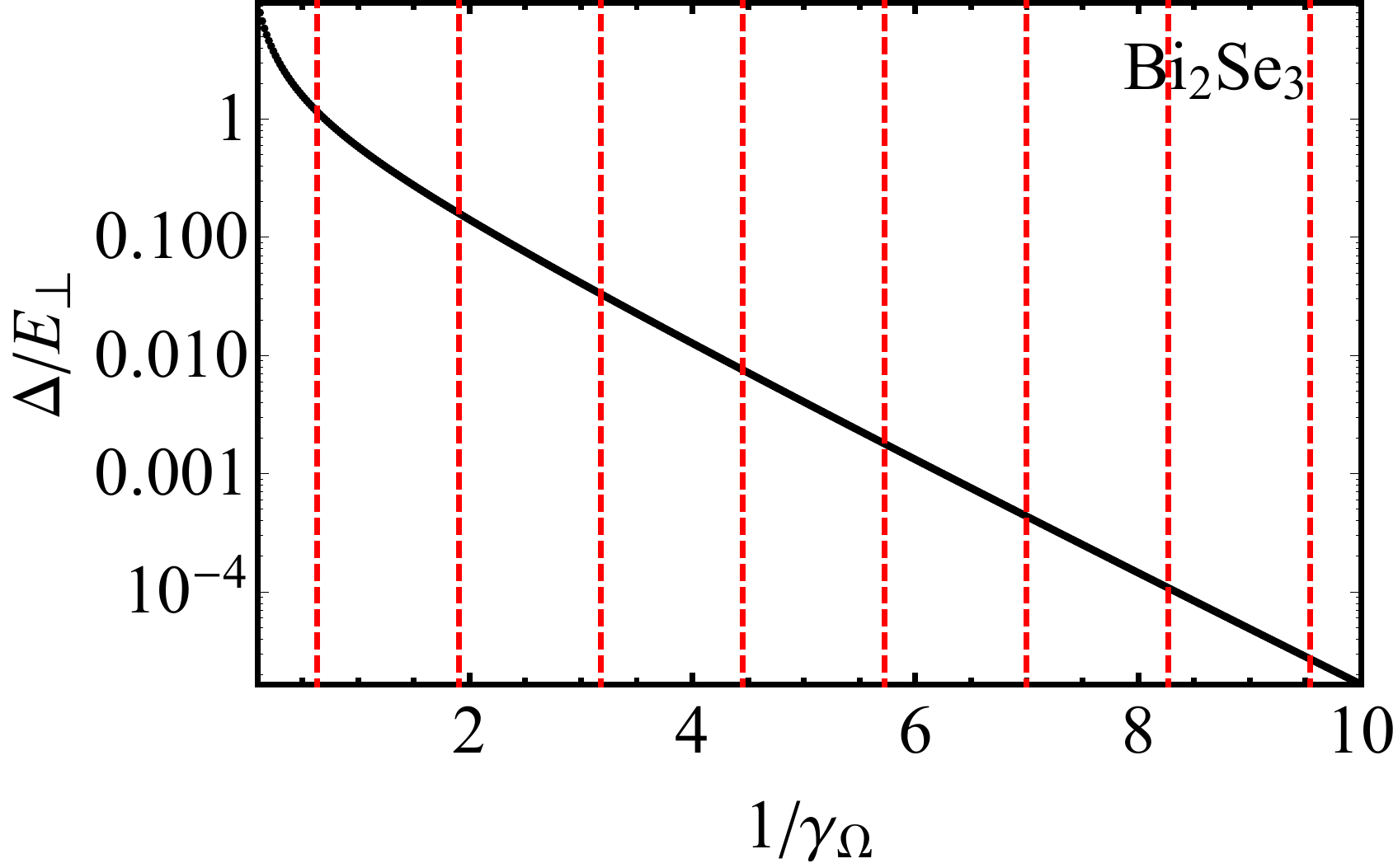}
\hspace{0.1cm}
\includegraphics[width=0.42\columnwidth]{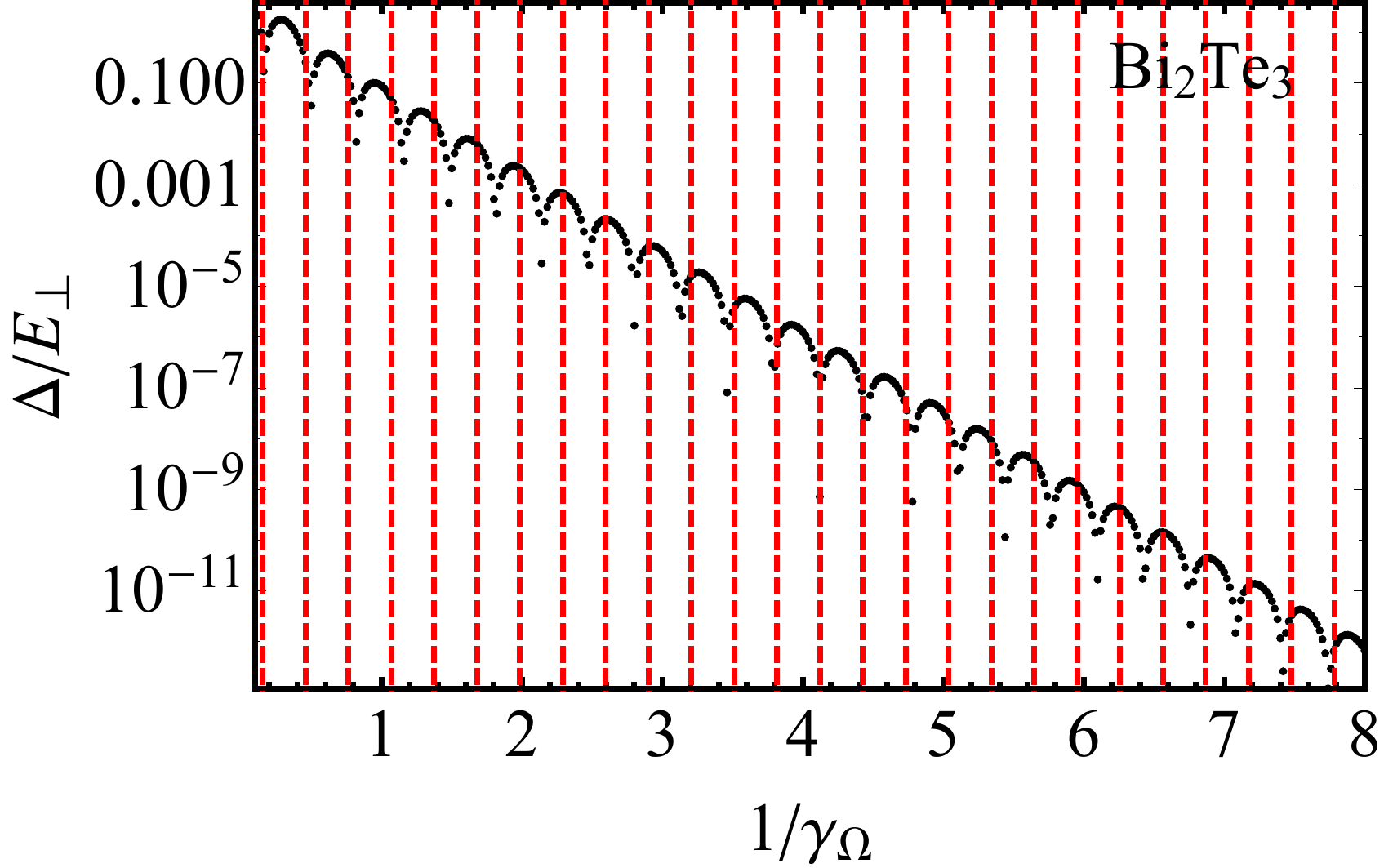}
\\[0.5cm]
\includegraphics[width=0.42\columnwidth]{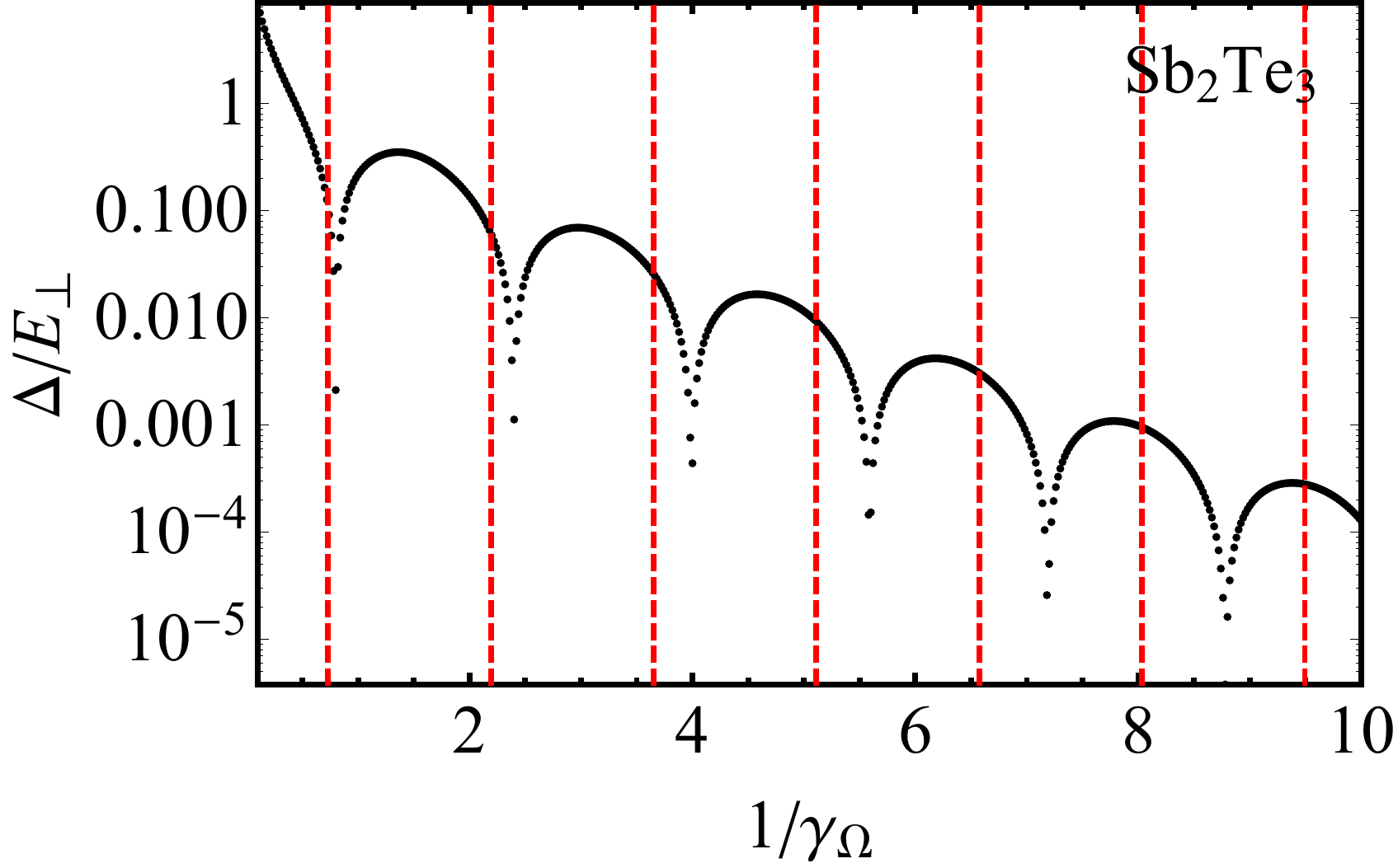}
\hspace{3.5cm}
\caption{Confinement dependence of the band gap $\Delta$, calculated with
$H_\epsilon$ neglected. The left (right, bottom) panel shows results obtained
using the materials parameters~\cite{nec16} for Bi$_2$Se$_3$ (Bi$_2$Te$_3$,
Sb$_2$Te$_3$), i.e., $\gamma_{M_0} = -0.785$ ($-3.27$, $-0.683$),
$\gamma_\parallel = 2.34$ ($0.484$, $0.117$). Vertical lines correspond to
values of $\gamma_\Omega$ for which the confinement-renormalized effective gap
$\Delta_{n\vek{0}}$ given in Eq.~(\ref{eq:topGap}) vanishes.}
\label{fig:bandgap}
\end{flushright}
\end{figure}

The fundamental gap $\Delta$ between the lowest conduction (sub-)band and the
highest valence (sub-)band is a quantity of both conceptual and practical
importance. It was shown to decrease exponentially as a function of system
size in the confined direction both in 2D~\cite{zho08} and 3D~\cite{lin09,
liu10a,lu10,sha10} TIs. For 3D TIs~\cite{lin09,oza14,bet16} and
inversion-symmetry-breaking 2D TIs~\cite{tak14}, the exponential decay of
$\Delta$ was observed to be modulated by an oscillation. Within our model, we
observe the oscillatory behavior, but with interesting materials-dependent
features. See Fig.~\ref{fig:bandgap}. The more strongly confinement-dominated
materials Bi$_2$Te$_3$ and Sb$_2$Te$_3$ (those with a rather small magnitude
of the parameter $\gamma_\parallel$) show clear oscillations, and the condition
for the bare subband crossings, i.e., vanishing effective Dirac gap given in
(\ref{eq:topGap}), quite accurately predicts the values of $\gamma_\Omega$
at which $\Delta$ is minimal. Thus the period of gap oscillations in these two
materials is consistent with $\delta (1/\gamma_\Omega) \sim 1/\gamma_{M_0}$,
which corresponds to a period $\delta l_\Omega \sim \sqrt{M_1/M_0}$ for
oscillation of the gap as a function of the effective 2D-system width scale
$l_\Omega$. This estimate agrees with similar ones derived previously using a
hard-wall confinement~\cite{liu10} and a tight-binding model~\cite{oza14},
respectively, but deviations from such effective-Hamiltonian-based results
occur in few-layer samples~\cite{for15,for16,nec16}. In contrast, no gap
oscillations are discernible in our model with parameters applicable to
Bi$_2$Se$_3$, for which the mixing of bare harmonic-oscillator subbands is
always important due to the large value of $\gamma_\parallel$. Hence, even
though transitions between normal and inverted phases occur typically for the
2D subbands in all three materials (see below), their heralding by gap
oscillations depends sensitively on band-structure details~\cite{oka14} and,
in contrast to expectation~\cite{oza14}, is therefore not universal. The
observed absence (presence) of gap oscillations in Bi$_2$Se$_3$ (Bi$_2$Te$_3$
and Sb$_2$Te$_3$) also suggests that the system is dominantly a strong (weak)
TI~\cite{imu12}.

\section{Confinement tuning of band inversion and surface-state hybridization}
\label{sec:bandinv}

The band inversion in TI materials can be measured directly by the pseudo-spin
projection of energy eigenstates. Within our present notation, ordinary
conduction (valence) band character of a state is quantified by the expectation
value $\braket{\tau_z\otimes\sigma_0}$ being close to $+1$ ($-1$). We use this
measure to study the fate of band inversion in quantum-confined TI systems,
complementing recent experimental studies~\cite{sal16}. Within our formalism
where a general state is given as a superposition
\begin{equation}
\Psi_{\kk_\perp}(z) = \sum_{n \alpha} \left[ b_{n\kk_\perp}^{(\alpha +)}\,
\psi_{n\kk_\perp\alpha}(z)\otimes\ket{+}_\mathrm{s} + b_{n\kk_\perp}^{(\alpha 
-)}\, \psi^\ast_{n\kk_\perp\alpha}(z)\otimes\ket{-}_\mathrm{s} \right] \, ,
\label{eq:totalwavefct}
\end{equation}
its pseudo-spin expectation value is straightforwardly calculated as
\begin{eqnarray}
\fl \braket{\tau_z\otimes\sigma_0}_{\kk_\perp} = \sum_{n \alpha \beta} \Bigg[
\sum_\tau \frac{\tau}{2} \left( 1 + \tau \, \frac{\Delta_{n\kk_\perp}}{E_{n
\kk_\perp \alpha}} \right)^{\frac{1}{2}} \left( 1 + \tau \, \frac{\Delta_{n
\kk_\perp}}{E_{n\kk_\perp \beta}} \right)^{\frac{1}{2}} \Bigg] \left[
\sum_\sigma \Re\mathrm{e} \left\{ \big[ b_{n\kk_\perp}^{(\alpha \sigma)}
\big]^* b_{n\kk_\perp}^{(\beta \sigma)} \right\} \right] . \nonumber \\
\end{eqnarray}

\begin{figure}
\begin{flushright}
\includegraphics[width=0.4\columnwidth]{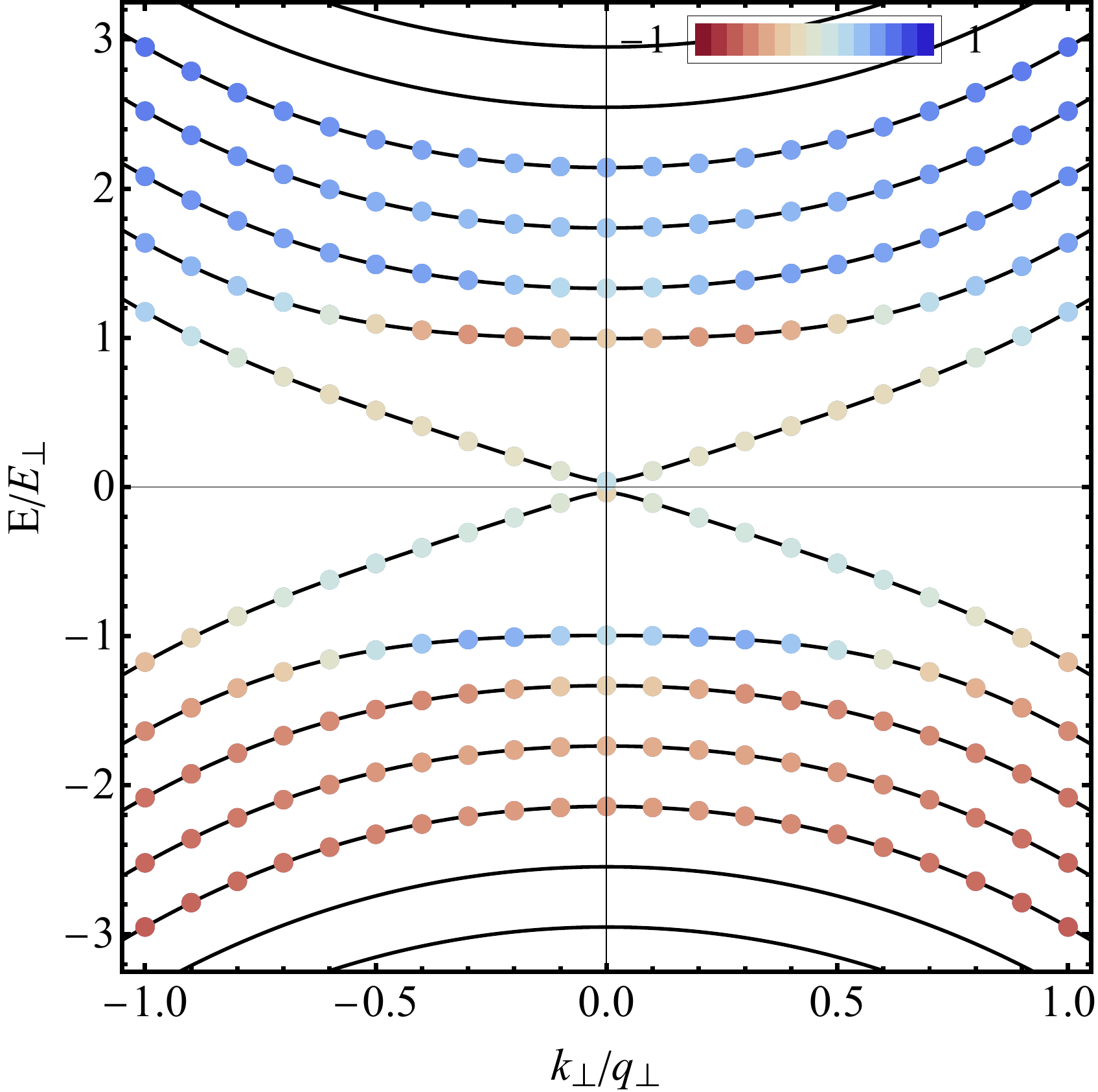}
\hspace{0.5cm}
\includegraphics[width=0.4\columnwidth]{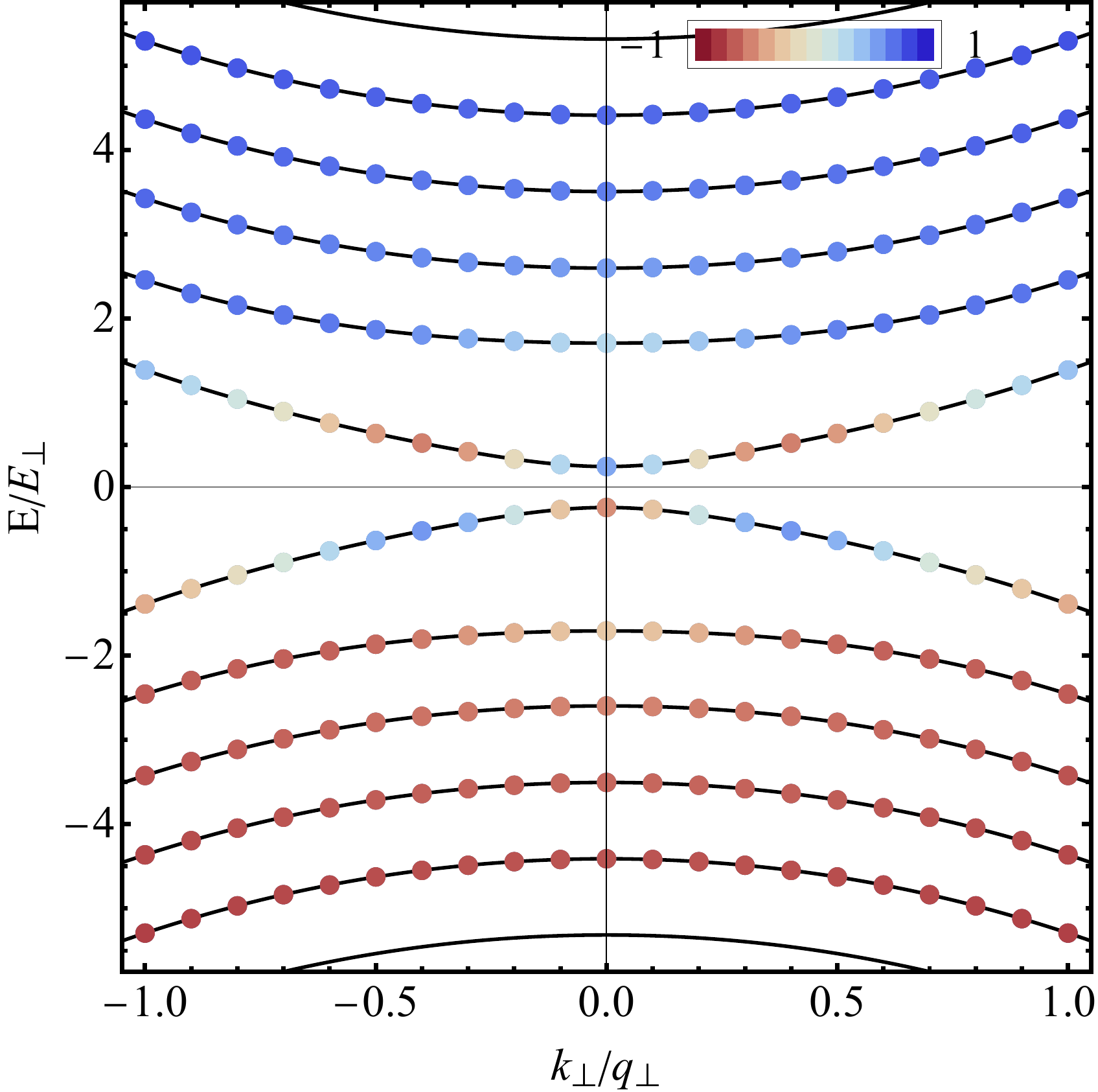}
\caption{Band inversion in a harmonically confined 3D topological insulator
with large band mixing ($\gamma_\parallel = 2.34$ as for Bi$_2$Se$_3$
\cite{nec16}) and electron-hole asymmetry neglected. The left (right) panel
corresponds to $\gamma_\Omega = 0.400$ ($0.900$). The color of a dot is a
measure for the pseudo-spin expectation value $\braket{\tau_z\otimes\sigma_0}$
of the corresponding eigenstate as per the given scale. Subbands at large
positive (negative) energies are found to have the ordinary
conduction-(valence-)band character, as signified by a pseudo-spin value close
to $+1$ ($-1$). Lower subbands in not-too-thin samples exhibit band inversion,
as exemplified by the 2nd subband in the left panel. The interplay of band
inversion and size quantization can also result in nontrivial $k_\perp$
dependences of the pseudo-spin projection, an example of which is shown by the
lowest subband in the right panel. See also the simulations illustrating the
evolution of subband structure and pseudo-spin character as the
confinement-related parameter $\gamma_\Omega$ is varied continuously, which
are provided as supplemental data.}
\label{fig:tau3profile}
\end{flushright}
\end{figure}

As the thickness of the TI layer decreases, more and more 2D subbands (those
with band index $n \gg n_\mathrm{c}$) show normal behavior, with conduction and
valence band states having the ordinary pseudo-spin character. Low-energy
subbands with $n \lesssim n_\mathrm{c}$ continue to have band inversion, but
the lowest-energy ones become increasingly dominated by confinement effects and
less by the band mixing arising from $H_\parallel$. In particular, the lowest
subband oscillates between normal and inverted character in the strongly
confinement-dominated regimes exhibited by Bi$_2$Te$_3$ and Sb$_2$Te$_3$.
Figure~\ref{fig:tau3profile} shows examples of such features using materials
parameters applicable to Bi$_2$Se$_3$. More extensive simulations provided as
supplementary data illustrate the evolution of band inversion as the
confinement-related parameter $\gamma_\Omega$ is varied continuously in the
three materials of interest. Generally, an oscillating change between normal
and inverted character of the lowest subband is observed to follow the cycle
of closings for the bare gap parameter $\Delta_{n\kk_\perp}$ from
Eq.~(\ref{eq:topGap}). To a lesser extent, this is the case for the
not-confinement-dominated Bi$_2$Se$_3$. Generally, the magnitude of the gap
appears to be more strongly affected by band mixing than the pseudo-spin
character of the size-quantized subband states. Thus it can be misleading to
use the measured sequence of gap minima to infer topological and normal regimes
in a confined system, whereas the pseudo-spin always serves as a reliable
identifier. The cycle of band inversions occurring in the bare
harmonic-oscillator part of the model embodied in $H_0(-i\partial_z, \kk_\perp)
+ H_V(z)$ generally coincides quite well with the succession of band inversions
for the lowest 2D subbands obtained from diagonalizing the full model $H_0
(-i\partial_z, \kk_\perp) + H_\parallel(-i\partial_z) + H_V(z)$.

Whether a 2D subband is inverted or normal is conventionally determined by the
character of states at the band edges. At finite in-plane wave vector
$\kk_\perp$, inter-band coupling due to the term proportional to $A_0$ in
(\ref{eq:inPlaneH}) induces a mixing of pseudo-spin states, and the term
proportional to $M_2$ contributes to the effective band gap. As a result, a
crossover from inverted to normal character generally occurs at large
$k_\perp$, as can be seen for the first and second subbands shown in the left
panel of Fig.~\ref{fig:tau3profile}. Interestingly, for low-lying 2D subbands
and in the confinement-dominated regime, band inversion can also just occur
within a region of finite $k_\perp$. See the right panel of
Fig.~\ref{fig:tau3profile} for a pertinent example, and the simulations
provided as supplemental data for further illustration. Again, it should be
emphasized that our results regarding the evolution of band inversions with
quantum confinement are based on the effective-Hamiltonian approach from which
we expect deviations to occur in few-layer samples~\cite{for15,for16,nec16}.

\begin{figure}
\begin{flushright}
\includegraphics[width=0.4\columnwidth]{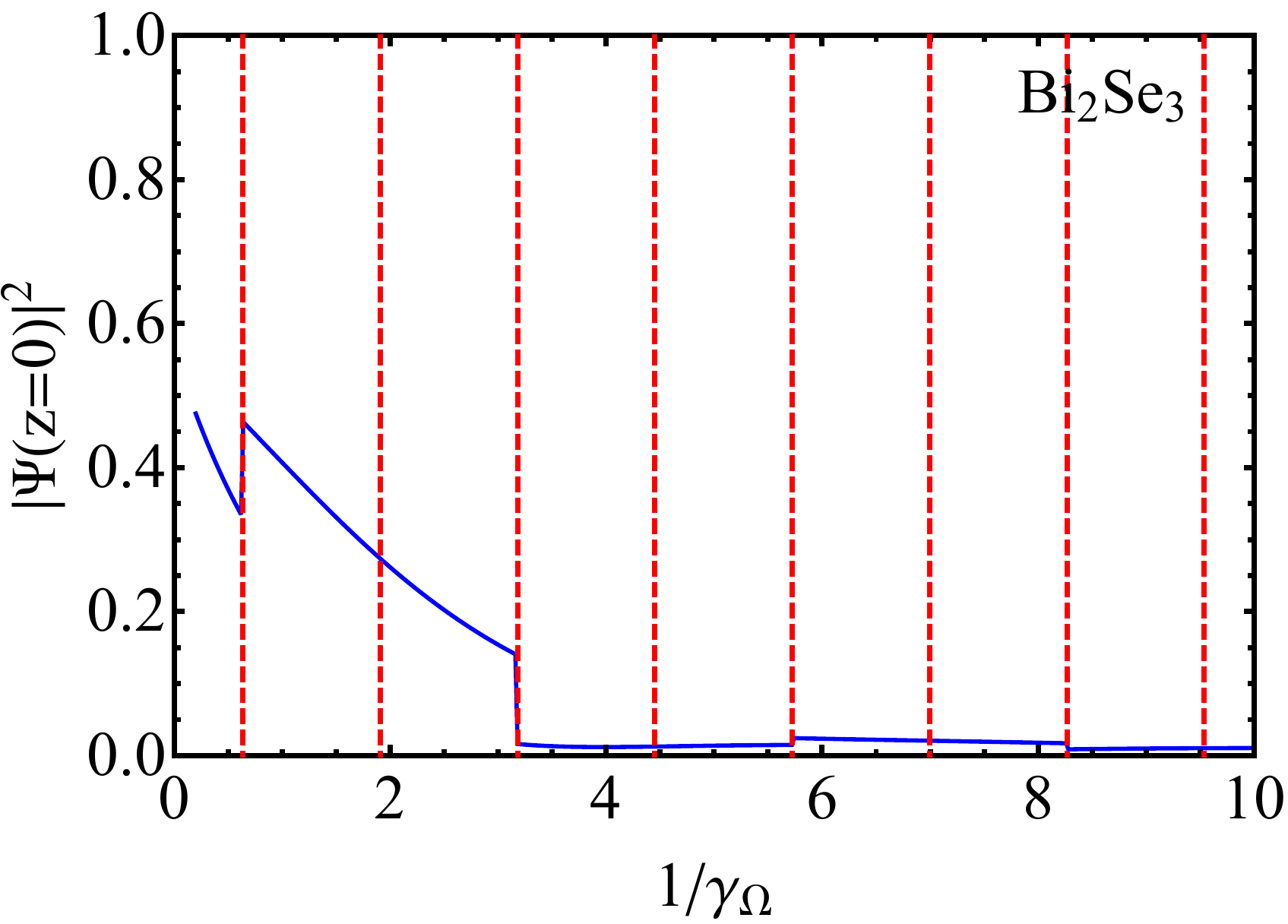}
\hspace{0.5cm}
\includegraphics[width=0.4\columnwidth]{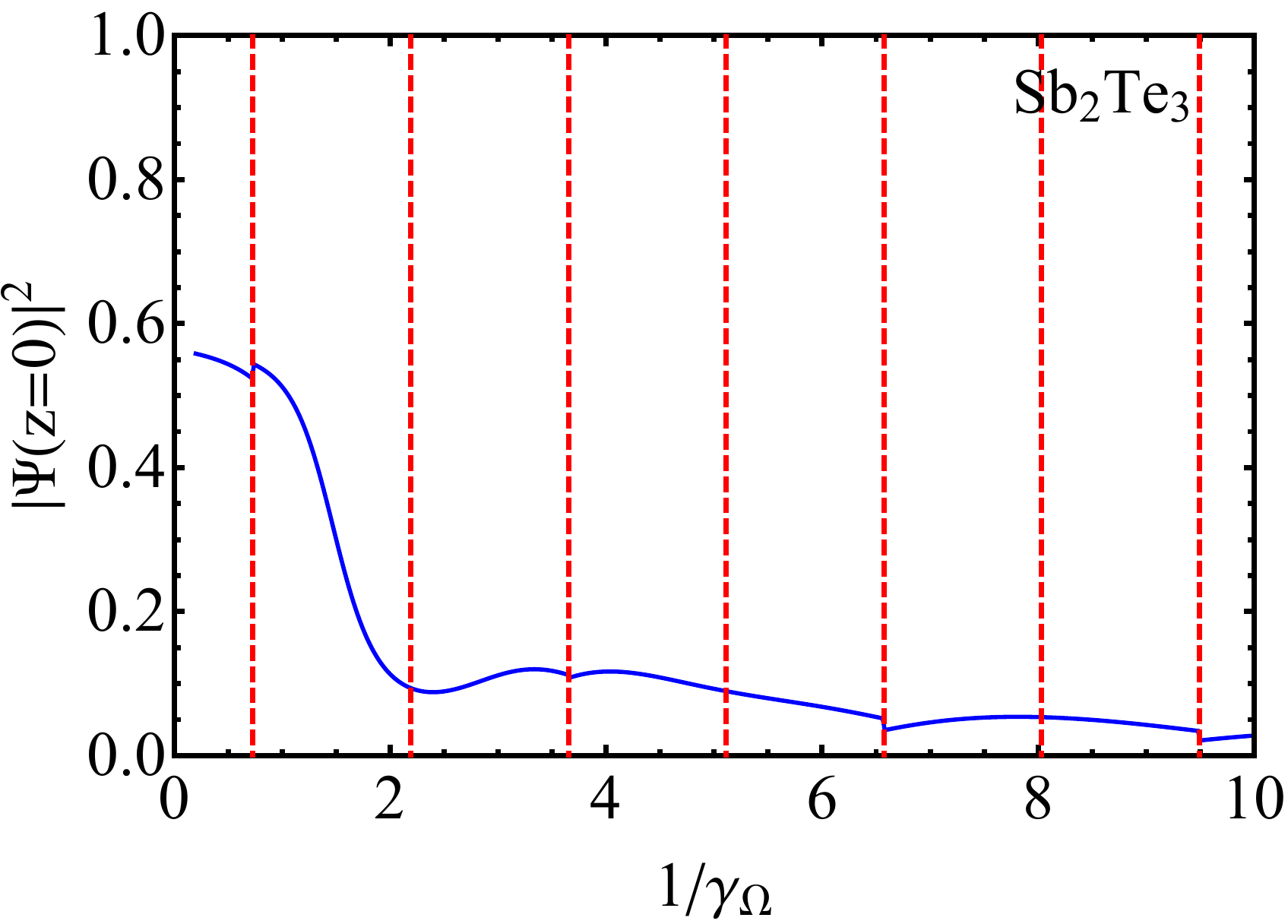}
\caption{Transition from boundary-localized to 2D-layer-extended character
of the lowest conduction subband-edge state, as measured by its probability
density $|\Psi_{\kk_\perp=0}(z)|^2$ at the 2D layer's center position $z=0$.
The left (right) panel shows results calculated for Bi$_2$Se$_3$
(Sb$_2$Te$_3$) materials parameters~\cite{nec16} with electron-hole asymmetry
neglected. Vertical lines correspond to values of $\gamma_\Omega$ for which
the confinement-renormalized effective gap $\Delta_{n\vek{0}}$ given in
Eq.~(\ref{eq:topGap}) vanishes.}
\label{fig:WFhybrid}
\end{flushright}
\end{figure}

In a thick sample, the lowest-lying subband corresponds to states localized at
the system's boundaries. As confinement is increased, i.e., sample thickness
reduced, hybridization of states from opposite surfaces becomes important, and
the states progressively loose their boundary-localized character. Eventually,
even the lowest-lying subband-edge states are extended over the entire
2D-layer width. This general behavior is instructively demonstrated by the
shape of probability-density distributions of lowest-subband-edge states shown
as part of the interactive simulations provided in the supplementary data. To
enable a more quantitative exploration of the transition between the
boundary-localized and 2D-layer-extended regimes, we introduce $l_\Omega\, 
|\Psi_{\kk_\perp}(z)|^2$ for $\kk_\perp=0$ and $z=0$ as a measure for the
hybridization-induced 2D-bulk character of subband-edge states. See
Fig.~\ref{fig:WFhybrid}. In Bi$_2$Se$_3$, small abrupt changes associated with
band inversions occur on top of a systematic change to the strongly hybridized
regime for $1/\gamma_\Omega \lesssim 3$. The practically vanishing
hybridization for a broad range of confinement strengths is indicative of
well-defined surface states. In contrast, the transition between
surface-localized lowest-energy states to 2D-bulk-like behavior is sharper in
the confinement-dominated case of Sb$_2$Te$_3$ but, at the same time,
hybridization remains substantial even for the larger-thickness samples.

\section{Effect of electron-hole asymmetry}\label{sec:ehAS}

So far, we have neglected the influence of asymmetries between conduction
and valence bands that are embodied in the contribution $H_\epsilon(k_z,
\kk_\perp)$ to the TI-material Hamiltonian given in (\ref{eq:phAsymmH}).
Comparing the magnitudes of $C_1$ and $C_2$ with those of $M_1$ and $M_2$,
respectively, in real materials would suggest that electron-hole asymmetry
is generally not a small effect. Here we investigate the latter's influence on
the modification of topological properties by quantum confinement.

\begin{figure}
\begin{flushright}
\includegraphics[width=0.4\columnwidth]{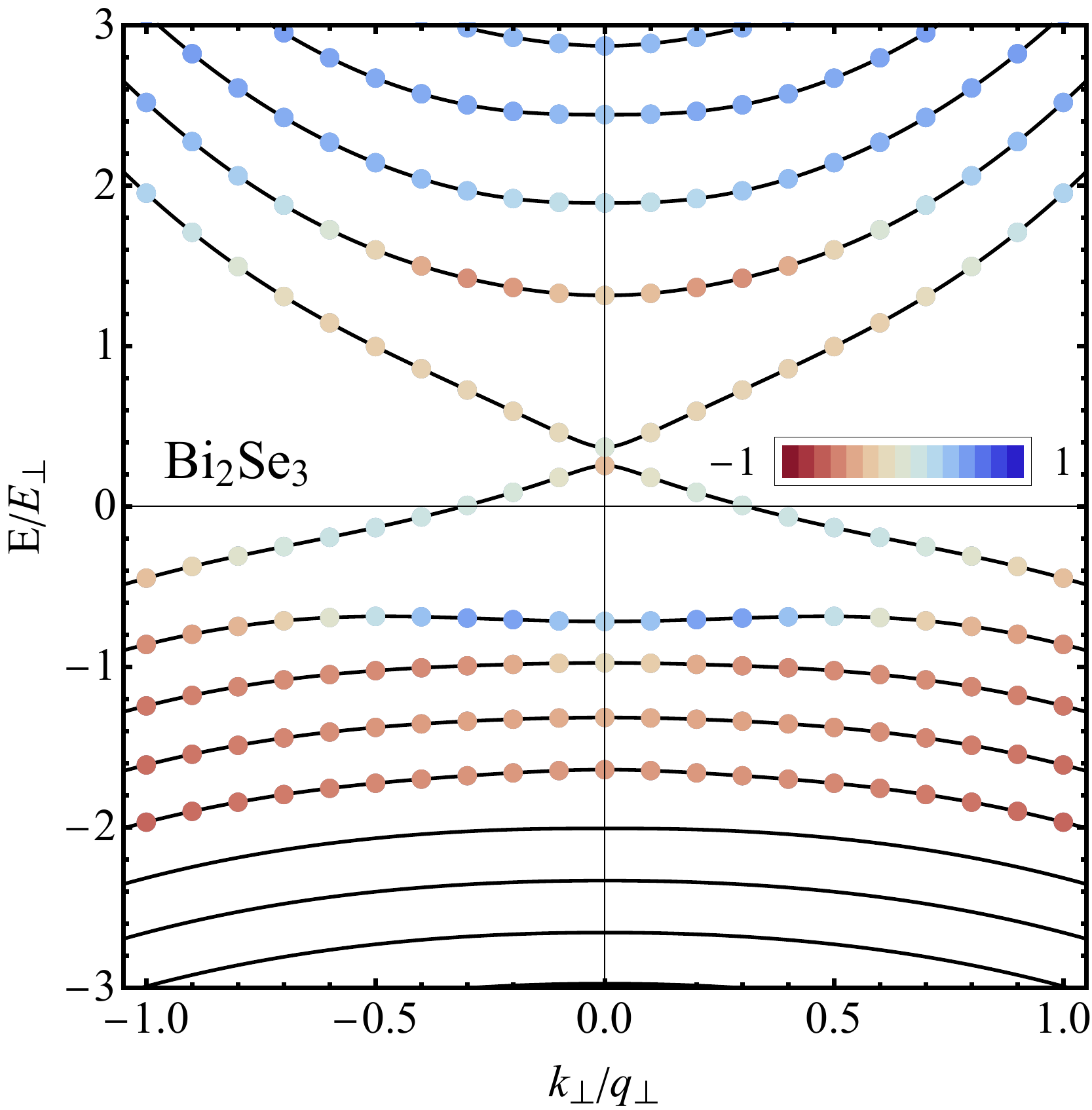}
\hspace{0.5cm}
\includegraphics[width=0.4\columnwidth]{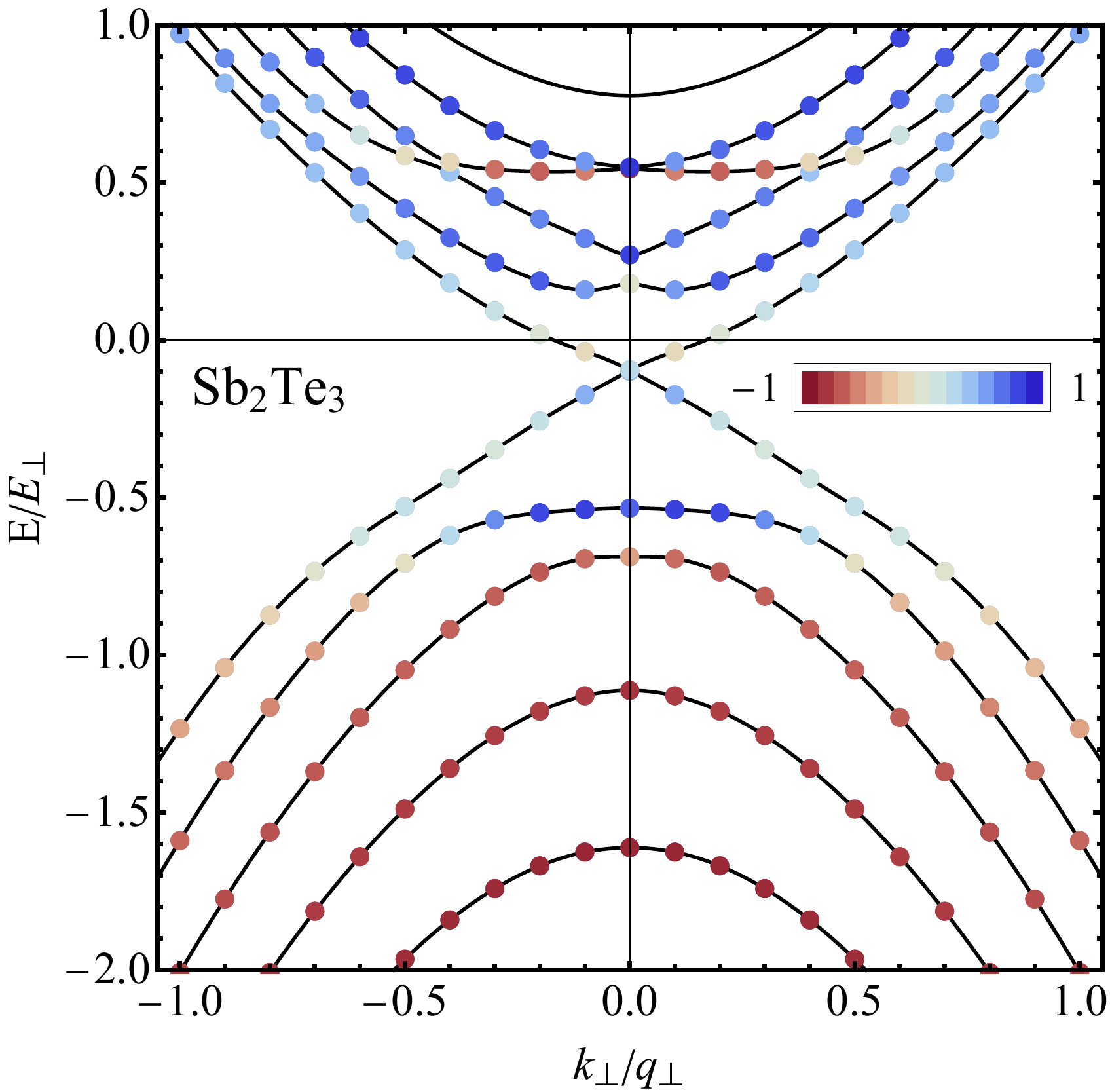}
\caption{Effect of electron-hole asymmetry on subband dispersions and the
pseudo-spin expectation value. The left (right) panel shows results calculated
with band-structure parameters~\cite{nec16} for Bi$_2$Se$_3$ (Sb$_2$Te$_3$)
and $\gamma_\Omega=0.400$.}
\label{fig:ehAsymm}
\end{flushright}
\end{figure}

The effect of the electron-hole-symmetry-breaking contribution $H_\epsilon(k_z,
\kk_\perp)$ on the subband dispersions generally turns out to be indeed quite
large, as a comparison of results corresponding to Bi$_2$Se$_3$ shown in the
left panels of Figs.~\ref{fig:tau3profile} and \ref{fig:ehAsymm} clearly
demonstrates. Most importantly, the position of the neutrality point for the
massless-Dirac-like surface states is shifted in energy, and the conduction and
valence subbands acquire different effective masses, even leading to a mass
inversion occurring in a valence subband. Significant qualitative differences
in the subband structure for Sb$_2$Te$_3$, shown in the right panel of
Fig.~\ref{fig:ehAsymm}, are a consequence of the different magnitudes and
opposite sign for the bulk-bandstructure parameters entering $H_\epsilon(k_z,
\kk_\perp)$.

Given the drastic changes exhibited in the lineshapes of subband spectra, one
may ask whether, and if so how, any of the above-discussed features associated
with the interplay of topological properties and quantum confinement survive in
the presence of electron-hole asymmetry. We focus here specifically on the
fundamental (i.e., lowest-subband energy) gap and the band inversion.
Interactive simulations are included as supplemental data to visualize the
situation. As it turns out, in the confinement-dominated regime, electron-hole
asymmetry is found to lead only to marginal quantitative changes. Basic
qualitative features such as gap oscillations and subband inversions occur very
similarly to the situation where $H_\epsilon(k_z,\kk_\perp)$ is neglected, with
the gap oscillations typically more significantly affected~\cite{oka14}. This is
particularly clearly exhibited by the case of Sb$_2$Te$_3$ for not-too-weak
confinement. But also for Bi$_2$Se$_3$ where mixing of bare-oscillator subbands
is strong, the evolution of band inversions is basically identical with and
without electron-hole asymmetry included, even for the lowest subband and
weakest confinement strengths. Interestingly, for the system whose parameters
correspond to Sb$_2$Te$_3$, the scrambling of low-energy subbands in the limit
of very weak confinement tends to obscure the topological-insulator properties
that are still unambiguously exhibited when electron-hole asymmetry is
neglected.

\section{Conclusions and outlook}\label{sec:concl}

We have used the effective-model description of bulk-TI band structures to
investigate the effect of a soft, harmonic-oscillator-type, quantum
confinement on physical properties that epitomize topological phases. Our
study focuses explicitly on three materials systems (Bi$_2$Se$_3$,
Bi$_2$Te$_3$ and Sb$_2$Te$_3$) that are representatives for quite different
regions of the 3D-TI parameter space and, hence, exhibit distinctive features
in the evolution of topological properties as the strength of the confinement
is varied. Interactive simulations have been included as supplemental data to
enable more detailed exploration of these differences. The results obtained
here are also straightforwardly generalized to any materials systems whose
band structure is described by the same type of $\kk\cdot\vek{p}$ Hamiltonian
that forms the basis for our theoretical approach, including Dirac
semimetals~\cite{wan13,xia15,pan15} and other surmised TIs such as
Bi$_2$Te$_2$I$_2$~\cite{nec16a}.

The interplay of band inversion, size quantization, and band mixing is found
to be governed by the relative magnitudes of unit-less parameters defined in
Eq.~(\ref{eq:parameters}). Characteristic features exhibited by the particular
materials systems considered here can thus be rationalized in terms of the
specific values of these parameters. Fundamentally, as the system size in the
confined direction varies, the gaps of inverted bare-oscillator subbands
[Eq.~(\ref{eq:topGap})] are successively closing and reopening. Mixing of
bare-oscillator subbands significantly modifies the bare-oscillator subband
dispersions in the large-width regime ($\gamma_\Omega < \gamma_\parallel$),
establishing the vanishing-mass Dirac-like surface-state dispersion and
eventually causing the disappearance of oscillations in the fundamental
(lowest-subband) gap value. We track the evolution of band inversions by
explicit consideration of the pseudo-spin value of eigenstates, establishing
the robustness of TI phases with respect to band mixing and electron-hole
asymmetry. In contrast, the occurrence of gap oscillations turns out to be a
nonuniversal feature that is generally not a useful measure to monitor TI
character in a quantum-confined system.

Our study considering 3D TIs subject to the harmonic-oscillator confining
potential complements previous works that assumed hard-wall confinement. In
contrast to the situation reported in single-particle analyses of 2D
TIs~\cite{sta09,sta10,buc12}, the basic features and overall trends associated
with the effects of size quantization on 3D-TI properties do not drastically
differ between the cases of hard-wall and soft-harmonic-oscillator potentials.
Instead of the type of confining potential, differences in the 3D-TI
bandstructure parameters are a major cause of significant variability in the
exhibited behaviors.

The conclusions presented in this work have been reached based on an approach
where certain aspects of real materials were ignored. One of these is the
finite size of thin-film samples in the plane perpendicular to the
quantum-confined direction and any effects arising from the conducting lateral
surfaces. However, apart from the edge-state structure in strong perpendicular
magnetic fields~\cite{liu16}, systems with a large-enough aspect ratio are
well-described by models assuming them to be infinite in the transverse
directions. Another potentially relevant issue concerns the interplay of
confinement and interactions. For example, the presence of long-range (Coulomb)
interactions is known to induce a reconstruction of the boundary-electronic
(edge-state) structure in 2D quantum-Hall~\cite{dem93,mac93,cha94,bar11} and
quantum-spin-Hall~\cite{wan17,ama17} systems. Although interaction effects
often turn out to be less pronounced in higher spatial dimensions, it would be
interesting for future studies to address the influence of Coulomb interactions
on harmonically confined 3D TIs, especially those where strong correlations are
expected to affect topological properties~\cite{hoh13,ama16}. The goal of such
investigations would be to establish how, and where in parameter space, the
purely single-particle-related confinement effects in 3D TIs discussed in the
present work are modified due to interactions.

\ack
Useful discussions with D~Abergel, P~Brydon, M~Governale, K~I~Imura, and H~Z~Lu
are gratefully acknowledged. We used a software tool available from
Ref.~\cite{mathematica-trick} to create the interactive graphics animations of
confined-3D-TI subband structures that are made available as supplementary data
in conjunction with this article.

\appendix
\section*{Appendix}
\setcounter{section}{1}

Here we provide some mathematical details for solving the Schr\"odinger
problem for a quantum-confined 3D TI.

When electron-hole asymmetry is neglected, the Schr\"odinger equation reads
\begin{equation}\label{eq:SEwoAS}
\left[ H_0 (-i\partial_z, \kk_\perp) + H_\parallel(-i\partial_z) + H_V(z)
\right] \Psi_{\kk_\perp}(z) = E_{\kk_\perp}\, \Psi_{\kk_\perp}(z) \quad .
\end{equation}
Making the \textit{Ansatz\/} (\ref{eq:totalwavefct}) transforms 
(\ref{eq:SEwoAS}) into a set of recursive relations
\begin{eqnarray}
\fl \frac{E_{m\kk_\perp\beta} - E_{\kk_\perp}}{E_\perp} \, b^{(\beta
\sigma)}_{m\kk_\perp} + \sigma \, i \sum_\alpha \Brac{C^{m-1,m}_{\alpha\beta,
\kk_\perp} \,\, b^{(\alpha,-\sigma)}_{m-1,\kk_\perp} + C^{m,m+1}_{\beta
\alpha, \kk_\perp} \, \, b^{(\alpha,-\sigma)}_{m+1,\kk_\perp}} = 0
\label{eq:recursiveeq}
\end{eqnarray}
for the unknown coefficients $b^{(\alpha\sigma)}_{n\kk_\perp}$, with the matrix
elements given explicitly by
\begin{eqnarray}
\fl C^{n,m}_{\alpha\beta,\kk_\perp} = \sqrt{\frac{m\, \gamma_\Omega \,
\gamma_\parallel}{8\, E_{n\kk_\perp\alpha} \, E_{m\kk_\perp\beta}}} \,
\Bigg[ \alpha\, \sqrt{\brac{E_{m\kk_\perp\beta} + \Delta_{m\kk_\perp}}
\brac{E_{n\kk_\perp\alpha} - \Delta_{n\kk_\perp}}} \nonumber \\ \hspace{4cm}
-\beta\, \sqrt{\brac{E_{m\kk_\perp\beta} - \Delta_{m\kk_\perp}}
\brac{E_{n\kk_\perp\alpha} + \Delta_{n\kk_\perp}}}\Bigg] \,\, .
\end{eqnarray}
We truncate this infinite set of equations by writing a matrix equation,
$\mathcal{H}_N\, b_N = E_N\, b_N$, with $N$ large enough as required to achieve
numerical accuracy, and the definitions
\begin{eqnarray}
b_N = \brac{b_{0,\kk_\perp}^{(++)}, b_{0,\kk_\perp}^{(-+)}, b_{0,
\kk_\perp}^{(+-)}, b_{0,\kk_\perp}^{(--)}, b_{1,\kk_\perp}^{(++)},\dots ,
b_{N -1, \kk_\perp}^{(--)}}^T \quad , \\[0.4cm]
\mathcal{H}_{N} =
\brac{\begin{array}{cccccccc} \mathcal{E}_0^\prime & \Gamma_0 &  &  &  &  &  &
\\ (\Gamma_0)^\dagger & \mathcal{E}_1^\prime & \Gamma_1 &  &  &  & 0 & \\ &
(\Gamma_1)^\dagger & \mathcal{E}_2^\prime & \Gamma_2 &  &  &  &  \\  &  & . & . 
& . &  &  &  \\  &  &  & . & . & . &  &  \\  & 0 &  &  & . & . & . &  \\  &  & 
&  &  & (\Gamma_{N-2})^\dagger & \mathcal{E}_{N-1}^\prime & \Gamma_{N-1} 
\end{array}} \label{eq:HamiltonApprox} \, , \\[0.4cm]
\mathcal{E}_n^\prime = \brac{\begin{array}{cccc} 
E_{n\kk_\perp +} & 0 & 0 & 0 \\ 0 & E_{n\kk_\perp -} & 0 & 0 \\ 0 & 0 & E_{n
\kk_\perp +} & 0 \\ 0 & 0 & 0 & E_{n\kk_\perp -} \end{array}} , \,
\Gamma_n = \brac{\begin{array}{cc} 
0 & \mathrm{i}C_n \\ -\mathrm{i} C_n & 0 \end{array}} , \\[0.4cm]
C_n = \brac{\begin{array}{cc}  
C^{n,n+1}_{++,\kk_\perp} & C^{n,n+1}_{+-,\kk_\perp} \\[0.3cm]
C^{n,n+1}_{-+,\kk_\perp} & C^{n,n+1}_{--,\kk_\perp} \end{array}}\quad .
\end{eqnarray}

For the case with electron-hole asymmetry included, the contribution
$H_\epsilon \brac{-i\partial_z,\kk_\perp}$ from (\ref{eq:phAsymmH}) needs to
be added to the Hamiltonian. As a result, the recursion relations 
(\ref{eq:recursiveeq}) are modified and now read
\begin{eqnarray}
\fl \Brac{\frac{E_{m\kk_\perp\beta} - E_{\kk_\perp}}{E_\perp} +
\gamma_\epsilon \,\gamma_\Omega \brac{m + \frac{1}{2}} + \frac{C_0}{E_\perp}
+ \frac{C_2}{M_2}\, \frac{\kk_\perp^2}{q_\perp^2}} \, b^{(\beta\sigma)}_{m
\kk_\perp} + \sum_\alpha \Big[ \sigma\, i\, C^{m-1,m}_{\alpha\beta, \kk_\perp}
\,\, b^{(\alpha,-\sigma)}_{m-1,\kk_\perp} \nonumber \\[0.2cm]
+ \,\, \sigma\, i\, C^{m,m+1}_{\beta \alpha, \kk_\perp} \,\,
b^{(\alpha,-\sigma)}_{m+1,\kk_\perp} + D^{m-2,m}_{\alpha\beta,\kk_\perp}\,\,
b^{(\alpha,\sigma)}_{m-2,\kk_\perp} + D^{m,m+2}_{\beta\alpha,\kk_\perp} \,\,
b^{(\alpha,\sigma)}_{m+2,\kk_\perp} \Big] = 0 \,\,\, ,
\end{eqnarray}
with $\gamma_\epsilon=C_1/(2 M_1)$ and 
\begin{eqnarray}
\fl D^{n,m}_{\alpha\beta,\kk_\perp}=\frac{\gamma_\epsilon\, \gamma_\Omega\,
\sqrt{m(n+1)}}{4\sqrt{E_{n\kk_\perp\alpha} E_{m\kk_\perp\beta}}} \Bigg[
\sqrt{\brac{E_{m\kk_\perp\beta} + \Delta_{m\kk_\perp}}\brac{E_{n\kk_\perp
\alpha} + \Delta_{n\kk_\perp}}} \nonumber \\[0.2cm] \hspace{3.5cm} +\,\,
\alpha\beta \, \sqrt{\brac{E_{m\kk_\perp\beta} - \Delta_{m\kk_\perp}}
\brac{E_{n\kk_\perp\alpha} - \Delta_{n\kk_\perp}}} \Bigg] \quad .
\end{eqnarray}
Consequently, this leads to an expression similar to (\ref{eq:HamiltonApprox}),
\begin{eqnarray}
\fl \mathcal{H}_{N}^\epsilon =
\brac{\begin{array}{ccccccccc} \mathcal{E}_0 & \Gamma_0 & \Delta_{0} &  &  &  & 
& & \\ (\Gamma_0)^\dagger & \mathcal{E}_1 & \Gamma_1 & \Delta_{1} &  &  &  & 0 
&\\ (\Delta_{0})^\dagger & (\Gamma_1)^\dagger & \mathcal{E}_2 & \Gamma_2 & 
\Delta_{2} &  &  & & \\  & . & . & . & . & . &  & & \\  &  & . & . & . & . & . 
& & \\  & 0 &  & . & . & . & . & . & \\  &  &  &  & (\Delta_{N-3})^\dagger & 
(\Gamma_{N-2})^\dagger & \mathcal{E}_{N-1} & \Gamma_{N-1} & \Delta_{N-1}
\end{array}}\,\, ,
\label{eq:HamiltonAllApprox}
\end{eqnarray}
with
\begin{eqnarray}
\mathcal{E}_n = \mathcal{E}_n^\prime + E_\perp \, \mathrm{diag}
\Brac{\gamma_\epsilon \, \gamma_\Omega\, \brac{n+\frac{1}{2}} +
\frac{C_0}{E_\perp} + \frac{C_2}{M_2} \, \frac{\kk_\perp^2}{q_\perp^2}}
\quad , \\[0.4cm]
\Delta_n = \brac{\begin{array}{cc} 
D_n & 0 \\ 0 & D_n \end{array}}\quad , \quad D_n = \brac{\begin{array}{cc}  
D_{++,\kk_\perp}^{n, n+2} & D_{+-,\kk_\perp}^{n, n+2} \\[0.3cm] D_{-+,
\kk_\perp}^{n, n+2} & D_{--,\kk_\perp}^{n, n+2} \end{array}} \quad .
\end{eqnarray}

\section*{References}


\providecommand{\newblock}{}

\end{document}